\documentclass{iaus}
\usepackage{graphicx}

\title{The SMC Super-Shells as Probes of the Turbulent Dynamics of the ISM}
\author{Itzhak Goldman$^{1, 2}$}
\affiliation{$^1$Department of Basic Sciences,
Afeka Tel Aviv Academic College of Engineering, Tel Aviv, Israel\\[\affilskip]
$^2$Department of Astronomy and Astrophysics, School of Physics and Astronomy, Tel Aviv University, Tel Aviv, Israel \break email: goldman@wise.tau.ac.il}

\pubyear{2006}
\volume{237}   
\pagerange{1 --3}
\date{?? and in revised form ??}
\jname{Triggered Star Formation in a Turbulent ISM}
\editors{B. G. Elmegreen \& J. Palous, eds.}

\begin{document}

\maketitle

\begin{abstract}

The spatial power spectrum of the HI 21 cm intensity  in the Small Magellanic Cloud (Stanimirovic  \etal 1999)  is a power law over scales as large as those of the SMC itself. It was interpreted as due to turbulence by Goldman (2000) and by Stanimirovic  \& Lazarian (2001).
The question is whether the power spectrum is indeed the result of a   dynamical turbulence or is merely the result of a structured static density.
In the turbulence  interpretation of Goldman (2000) the turbulence was generated by the tidal effects of the last close passage of the LMC about 0.2 Gyr ago. The turbulence time-scale was estimated by Goldman to be 0.4 Gyr, so the turbulence has not decayed yet.
Staveley-Smith \etal (1997) observed in the SMC about five hundreds of  HI super shells. Their age is more than an order of magnitude smaller than   the turbulence age. Therefore, if the turbulence explanation holds, their observed radial velocities should reflect the turbulence in the gas in which they formed. 
In the present work we analyze the  observed radial velocities of the super shells. We find that the velocities indeed manifest  the statistical spatial correlations expected from turbulence. The turbulence spectrum is consistent with that obtained by Goldman(2000). 
\end{abstract}

\section{Introduction}

The spatial power spectrum of the HI  21 cm  intensity  in the Small Magellanic Cloud  was obtained by Stanimirovic  \etal (1999). Interestingly, it  is a power law over scales as large as that of the SMC itself. Similar power laws have been observed by  Crovisier \& Dickey
(1983) and by Green (1993) in the galaxy. The outstanding feature in the case of the SMC is the large scale of the observed correlations.
The power laws signal underlying long range correlations in what looks like
a field of random fluctuations of the intensity.
For an optically thin medium along the line of sight, the 21 cm intensity is proportional to the column density. Therefore, the fluctuations in 21 cm intensity, represent fluctuations in density.

A natural interpretation of the observed power spectra is that the underlying correlations in density fluctuations are due to a turbulence in which velocity fluctuations, that are coupled to density fluctuations, give rise to the observed power laws.  The turbulence interpretation was suggested by Goldman (2000) and Stanimirovic  \& Lazarian (2001).  

Goldman (2000)  suggested that this large scale turbulence was
generated by instabilities in the  bulk flows that resulted from the 
tidal interaction during the last close passage of  the Large
Magellanic Cloud (LMC) $\sim 2 Gyr$ ago (Gardiner \& Noguchi 1996).

However, since the observations catch a snapshot of the intensity field and since the turbulence timescales are very long ($\sim 0.4 $Gyr) one  cannot rule out the possibility of
 a {\it static} correlated density field that reflects  initial conditions.

In the present paper we propose a test to decide between these two alternatives.

\section{Analysis of the super shells radial velocity field} 

Staveley-Smith \etal (1997) observed 501 HI super shells in the SMC.
The proposed test relies on the fact that the timescale and age of the turbulence (if indeed there)
are typically $1 \div 2$ orders of magnitude larger than the lifetimes of the super shells. Therefore, they have formed in the turbulent gas and their observed radial velocities should reflect the turbulent velocity field in the gas in which they where formed. We wish to look at them as markers registering the ambient gas velocity. If the radial velocity field  exhibits spatial correlations consistent with the 
those of the turbulence, assumed as responsible for the 21 cm intensity spectra, it will strengthen the case for dynamical turbulence as the source of the HI intensity power spectrum.

We use the data of the 501 super shells reported in 
  Table 1 of Staveley-Smith \etal (1997). 
For each super shell, the residual radial velocity was found by subtracting from the observed velocity the large scale best fit, up to a   shear.

\begin{equation}
\label{resdv}
v_i=v_{obs, i}  - (c +s_1 x_i + s_2 y_i)  
\end{equation}
with $1\leq i \leq 501$
where
$$c=155.1 km/s, \ \ \ \ s_1=12.34 \times 10^{-3} km/s/pc, \ \ \ \ s_2=4.46\times 10^{-3} km/s/pc$$
 
The coordinates of each shell $(x_i, y_i) $ are in units of pc and were obtained from the angular coordinates by
adapting a distance of 60 kpc to the SMC. The velocities are in units of km/s. The subtracted large scale   velocity field 
is composed of a mean velocity and a shear. The magnitude of the shear is consistent
with values obtained by Gardiner  \& Noguchi (1996). 

 We have computed the second order structure function and the autocorrelation for the residual velocity field along lines parallel to the coordinate axes. Interpolation was used to fit the discrete data along the lines to a continuous function.
The different lines yielded similar results.

For simplicity, homogeneous and isotropic velocity field is assumed. In this case, the structure function and the autocorrelation depend only on the distance between the two points, $r=|\vec{r}|$. The structure function is

\begin{equation}
\label{struct} 
S(r)=<\left(v(\vec{r'} +\vec{r})-v(\vec{r'})\right)^2>  
\end{equation}
Similarly, the autocorrelation function is 

\begin{equation}
\label{cor} 
C(r)=< v(\vec{r'} +\vec{r}) v(\vec{r'})>  
\end{equation}
The angular brackets denote ensemble averaging. Assuming ergodicity, in addition to homogeneity and isotropy,  ensemble averaging equals space averaging.  As stated above, we use averages over lines so that

\begin{equation}
\label{sl} 
S(l)=\frac{1}{L}\int _0^L \left(v(x+l) -v(x)\right)^2 dx
\end{equation}
where $L$ is the length of the line. Similarly,
\begin{equation}
\label{cl} 
C(l)=\frac{1}{L}\int _0^L v(x+l) v(x) dx
\end{equation}
 The results of a typical computation are presented in figures 1-2. Figure 1 shows the structure function $S(l)$. For very small values of $l$ $S(l)\propto l^2$, for larger values of $l$ it varies as
$S(l)\propto l^{m-1}$ and then it saturates.

The index $m$  characterizes the inertial range of the  turbulent velocity spectral function: $F(k)\propto k^{-m}$. In Kolmogorov turbulence characterizing incompressible fluid $m=5/3$. In the case of turbulence in compressible medium $m=2$. This was also the value deduced by Goldman (2000) on the basis of the 21 cm intensity power spectrum.
These two power lows are presented in figure 1. The precision of the data is not enough to decide
between them, even though the $m=2$ line seems to follow better the slope of the
computed structure function.

\begin{figure}
\centerline{
\includegraphics[]{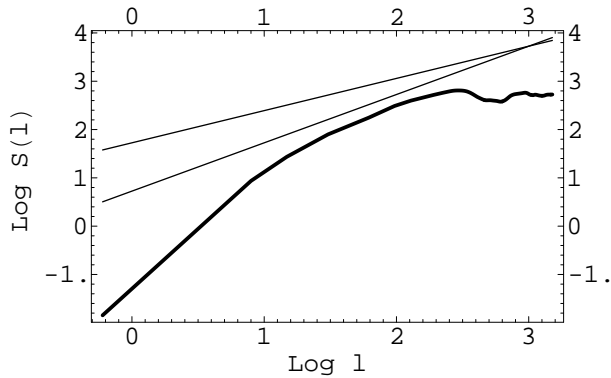}}
\caption{The structure function in units of $(km/s)^2$ as function of scale in pc. The thin lines  have slopes $2/3$ and  $1$. The upper line has a slope of $2/3$.} 
\end{figure}
The autocorrelation function is shown in figure 2. It behaves as  an autocorrelation function 
of a turbulent velocity rather than uncorrelated velocity fluctuations.

\begin{figure}
\centerline{
\includegraphics[]{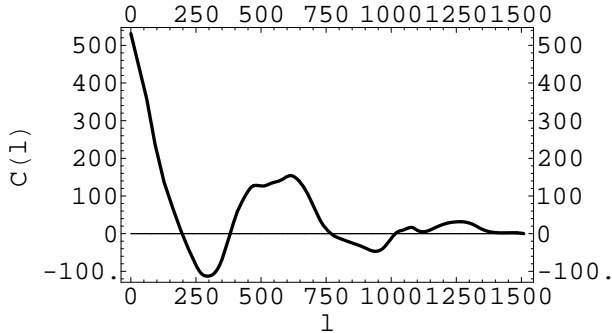}}
\caption{The  autocorrelation function in units of $(km/s)^2$ as function of scale in pc.}
\end{figure}

Figure 3 presents the turbulence spectral function $F(k)$ computed from the autocorrelation function. The curve is noisy but a power law range is clear. Also here the turbulence spectral functions with $m=5/3$ and $m=2$ are plotted. The two slopes are compatible with the computed 
spectral function, although $m=2$ seems preferable. 

The wavenumber range shown corresponds to spatial scales between $1500 pc$, which is in this case the length of the line $L$, and $50 pc$. Higher wavenumbers correspond to   spatial scales   that are smaller than the average radius of the shells, and therefore the  computed turbulence spectrum is 
not valid for these scales.

\section{Conclusions}
The results of the present work strengthen the case for the turbulence interpretation
of the 21 cm power spectra of the SMC. The residual radial velocities of the super shells exhibit statistical spatial correlations expected from turbulence. The turbulence spectrum and structure function are consistent     
with a Kolmogorov spectrum, $m =5/3$, and with that of incompressible turbulence, $m=2$. The
latter seems preferable. It equals the value deduced by Goldman (2000) from the HI intensity fluctuations.

\begin{figure}
\centerline{
\includegraphics[]{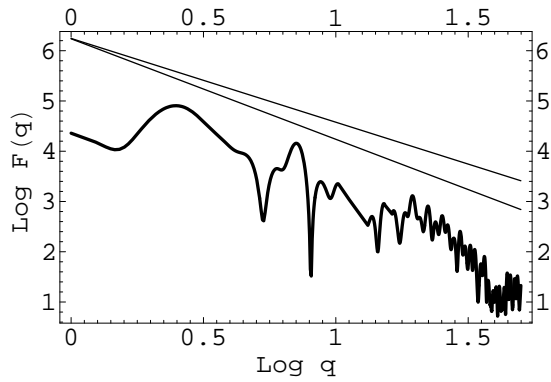}}
\caption{The turbulence spectral function in units of $(km/s)^2 pc$ as function of the normalized wavenumber  $q=k L/(2\pi)$. The  upper thin line   has a slope  $-5/3$ and the lower $-2$.}
\end{figure}

\begin{acknowledgments}
Participation in IAU Symposium 237 was    supported by Afeka Engineering College and by the Institute of Astronomy,
Department of Astronomy and Astrophysics, Tel Aviv University.  
\end{acknowledgments}

{}
\end{document}